\documentclass[preprint,showkeys,secnumarabic,amsfonts,showpacs,amsmath,amssymb]{revtex4}
\usepackage[dvips]{color}
\usepackage{array}
\usepackage{rotating}
\usepackage{epsfig}
\usepackage{amsmath}
\usepackage{graphicx}
\begin{document}
\title{The generalized second law in chameleon cosmology }

\author{H. Farajollahi}
\email{hosseinf@guilan.ac.ir} \affiliation{Department of Physics,
University of Guilan, Rasht, Iran}
\author{A. Salehi}
\email{a.salehi@guilan.ac.ir} \affiliation{Department of Physics,
University of Guilan, Rasht, Iran}
\author{F. Tayebi}
\email{ftayebi@guilan.ac.ir} \affiliation{Department of Physics,
University of Guilan, Rasht, Iran}
\date{\today}

\begin{abstract}
 \noindent \hspace{0.35cm}

In this paper, we investigate the validity of the generalized second law (GSL) of thermodynamics
 in flat FRW chameleon cosmology where the boundary of the
universe is assumed to be enclosed by the dynamical apparent horizon. It has been shown that, in a bouncing scenario for the universe with phantom crossing, the total entropy decreases with time in the contracting epoch, whereas, the dynamics of the internal and horizon entropies depends on the behavior of both  equation of state and hubble parameters.

\end{abstract}


\maketitle

\section{introduction}

Motivated by the black hole physics, it was realized that there is a profound connection
between gravity and thermodynamics. In Einstein gravity, the evidence of this connection
was first discovered in {\cite{Rindler}} by deriving the Einstein equation from the proportionality
of entropy and horizon area together with the first law of thermodynamics in the Rindler
spacetime. For a general static spherically symmetric spacetime, Padmanabhan pointed out
that Einstein equations at the horizon give rise to the first law of thermodynamics {\cite{first law}}. Recently,
the study on the connection between gravity and thermodynamics has been extended
to cosmological context. Frolov and Kofman {\cite{Frolov}} employed the approach proposed by Jacobson
{\cite{Rindler}} to a quasi-de Sitter geometry of inflationary universe, and calculated the energy flux
of a background slow-roll scalar through the quasi-de Sitter apparent horizon. By applying
the first law of thermodynamics to a cosmological horizon, Danielsson obtained Friedmann
equation in the expanding universe {\cite{Friedmann eq.}}. In the quintessence dominated accelerating universe,
Bousso {\cite{apparent horizon}} showed that the first law of thermodynamics holds at the apparent horizon. The
relation between gravity and thermodynamics has been further disclosed in extended gravitational
theories, including Lovelock gravity \cite{extended}--\cite{Lovelock gravity}, braneworld gravity \cite{braneworld gravity}--\cite{gravity}, nonlinear gravity
\cite{Lovelock gravity}, \cite{nonlinear gravity}--\cite{scalar-tensor gravity}, scalar-tensor gravity \cite{Lovelock gravity}--\cite{scalar-tensor gravity}, etc. In scalar-tensor
gravity, it was argued that the non-equilibrium thermodynamics instead of the equilibrium
thermodynamics should be taken into account to build the relation to gravity \cite{nonlinear gravity}--\cite{scalar-tensor gravity}. Determining thermodynamic parameters of an (accelerated) expanding
universe and verifying the first and the second thermodynamics law for different cosmological
horizons {\cite{thermo for different cosmological horizons}} and investigating the relation between dynamics and thermodynamics
of the universe {\cite{relation between dynamics and thermodynamics}} have also been the subjects
of many researches in recent years. In particular, the validity of the generalized second law (GSL) {\cite{GSL}} which state that entropy of the fluid
inside the horizon plus the entropy associated with the apparent horizon do not decrease
with time, has been the subject of many studies. In this paper, we study the thermodynamics properties of the chameleon cosmology in which the universe undergoes a kind of bouncing phase transition and phantom crossing. In section two, we drive the field equations for the model. The thermodynamics properties of the model is also studied in section three.

\section{The Model}

In this section, we consider the chameleon gravity in the presence of matter with the action given by \cite{farajollahi},
\begin{eqnarray}\label{action}
S=\int[\frac{R}{16\pi
G}-\frac{1}{2}\phi_{,\mu}\phi^{,\mu}+V(\phi)+f(\phi)L_{m}]\sqrt{-g}dx^{4},
\end{eqnarray}
where $R$ is Ricci scalar, $G$ is the newtonian constant gravity
and $\phi$ is the chameleon scalar field with the potential
$V(\phi)$. Unlike the usual Einstein-Hilbert action, the matter
Lagrangian $L_{m}$ is modified as $f(\phi)L_{m}$, where $f(\phi)$ is
an analytic function of $\phi$. This last term in Lagrangian brings about the nonminimal
interaction between matter and chameleon field. The variation of the action (\ref{action}) with respect to the metric tensor components in a spatially flat FRW  cosmology yields the field equations,
\begin{eqnarray}
&&3H^{2}=\rho_{m}f+\frac{1}{2}\dot{\phi}^{2}+V(\phi),\label{fried1}\\
&&2\dot{H}+3H^2=-\gamma\rho_{m}f-\frac{1}{2}\dot{\phi}^{2}+V(\phi),\label{fried2}
\end{eqnarray}
where we put  $8\pi G=c=\hbar=1$ and assume a perfect fluid with $p_{m}=\gamma\rho_{m}$. The dots means derivatives with respect to the cosmic time $t$. The energy density $\rho_{m}$ is the matter energy density in the universe. Also
variation of the action (\ref{action}) with respect to the scalar field  $\phi$ provides the wave
equation for the chameleon field as,
\begin{eqnarray}\label{phiequation}
\dot{\phi}(\ddot{\phi}+3H\dot{\phi})=-\dot{V}-\frac{1}{4}(1-3\gamma)\rho_{m}\dot{f}.
\end{eqnarray}
From equations (\ref{fried1}), (\ref{fried2}) and (\ref{phiequation}), one can easily arrive at the relation,
\begin{eqnarray}\label{conserv}
\dot{(\rho_{m}f)}+3H(1+\gamma)\rho_{m}f=\frac{1}{4}\rho_{m}(1-3\gamma)\dot{\phi}\dot{f},
\end{eqnarray}
which readily integrates to yield
\begin{eqnarray}
\rho_{m}=\frac{M}{f^{\frac{3}{4}(1+\gamma)}a^{3(1+\gamma)}},
\end{eqnarray}
with $M$ as a constant of integration. From equations (\ref{fried1}) and (\ref{fried2}) and in comparison with the standard friedmann equations we identify $\rho_{eff}$ and $p_{eff}$ as,
\begin{eqnarray}\label{roef}
\rho_{eff}\equiv\rho_{m}f+\frac{1}{2}\dot{\phi}^{2}+V(\phi).
\end{eqnarray}
\begin{eqnarray}\label{pef}
p_{eff}\equiv\gamma\rho_{m}f+\frac{1}{2}\dot{\phi}^{2}-V(\phi),
\end{eqnarray}
with the equation of state, $p_{eff}=\omega_{eff}\rho_{eff}$. From equations (\ref{roef}) and (\ref{pef}), one leads to,
\begin{eqnarray}\label{faydot}
\dot{\phi}^2=\frac{2}{\omega_{eff}-1}[\rho_{m}f(\gamma-\omega_{eff})-V(1+\omega_{eff})].
\end{eqnarray}
In the next section we investigate the thermodynamics and GSL of the model.

\section{GSL }

In the following we make two assumptions: 1) In addition to the entropy of the universe inside the horizon, an entropy is associated to the apparent/event horizon. 2) with the local equilibrium hypothesis, the
energy would not spontaneously flow between the horizon and the fluid inside the horizon, the latter would be
at variance with the FRW geometry.

According to the GSL in an expanding universe, entropy of the viscous DE, DM and radiation
inside the horizon plus the one associated with the apparent horizon do not decrease
with time. However, in the cosmological models where the universe has a bouncing behavior, the rate of change of the entropy can be negative as will be discussed shortly.

In general, there are two approaches to validate the GSL on apparent/event horizons: i) by using first law of thermodynamics and find entropy relation on the horizons i.e., \cite{horizon.first-law1},\cite{horizon.first-law.Bousso},
\begin{eqnarray}\label{first-law1}
T_{h}dS_{h}=-dE_{h}=4 \pi R_{h}^{3} H T_{\mu\nu}\kappa^{\mu}\kappa^{\nu}dt=4 \pi R_{h}^{3} H (\rho_{eff}+p_{eff})dt,
\end{eqnarray}
where $\kappa^{\mu} = (1, -Hr, 0, 0)$ are the (approximate) Killing vector ( the generators of the horizon), or the
future directed ingoing null vector field \cite{horizon.first-law.Wang} and $"h"$ stands for the horizon. ii) in the field equations, by employing the horizon entropy and temperature formula on the horizon \cite{GSL in BD},
\begin{eqnarray}
S_{h}=\pi R_{h}^{2},\label{entropy}\\
T_{h}=\frac{1}{2\pi R_{h}}\label{temp.1}.
\end{eqnarray}
Note that only on the apparent horizon the two approaches are equivalent \cite{GSL in BD}. Furthermore, the recent observational data from type Ia Supernovae suggests that in an accelerating universe the enveloping surface should be the apparent
horizon rather than the event one \cite{AH instead EH}. Therefore, from now on, we assume that the universe is enclosed by the dynamical apparent horizon with the radius given by $R_{h}=\frac{1}{\sqrt{H^{2}+\frac{k}{a^2}}}$ \cite{R_AH}.

By the horizon entropy and temperature given in equations (\ref{entropy}) and (\ref{temp.1}),
the dynamics of the entropy on the apparent horizon is \cite{GSL},
\begin{eqnarray}\label{entropyAH}
\dot{S_{h}}=2\pi R_{h} \dot{R_{h}}.
\end{eqnarray}
Also, from the Gibbs equation, the entropy of the universe inside the horizon can be related
to its effective energy density and pressure in the horizon with \cite{Gibb's eq.}
\begin{eqnarray}\label{Gibb's eq.1}
TdS_{in}=p_{eff}dV+d(E_{in}),
\end{eqnarray}
where $S_{in}$ is the internal entropy within the apparent horizon and $p_{eff}$ is the effective pressure in the model. If there is no energy exchange between outside and inside of the apparent horizon, thermal equilibrium realizes that $T = T_{h}$. Here, the expression for internal energy can be written as $E_{in} = \rho_{eff}V$, with $V = \frac{4}{3}\pi R_{h}^{3}$. Therefore, from equation (\ref{Gibb's eq.1}), by using Friedmann equation and after doing some calculation we find that the rate of change of the internal entropy, horizon entropy and total entropy are respectively,
\begin{eqnarray}
\dot{S_{in}}&=&12\pi^2 R_h^2H(1+\omega_{eff})(1+3\omega_{eff}),\label{sdotin}\\
\dot{S_{h}}&=&24\pi^2 R_h^2H(1+\omega_{eff}),\label{sdoth}\\
\dot{S_{t}}&=&36\pi^2 R_h^2H(1+\omega_{eff})^2.\label{sdott}
\end{eqnarray}
As already noted the GSL states that the total entropy is not a decreasing function of time, or $\dot{S_{t}}\geq 0$. However, if we assume a bouncing universe as in our chameleonic model, then we expect a decreasing total entropy when the universe is in contracting epoch.

For the rate of change of the internal entropy, equation (\ref{sdotin}), we find that when $H<0$ and $\omega_{eff}\leq -1$ or $\omega_{eff}\geq-1/3$, then $\dot{S_{in}}\leq 0$. Also, when $H>0$ and $-1 <\omega_{eff}<-1/3$, again $\dot{S_{in}}< 0$. On the other hand, when $H<0$ and $-1\leq\omega_{eff}\leq-1/3$ or $H>0$ and $\omega_{eff}<-1$ or $\omega_{eff}>-1/3$, we obtain that $\dot{S_{in}}>0$. we conclude that the behavior of the internal entropy depends both on the hubble parameter and effective EoS parameter of the model. From equation (\ref{sdoth}), it can be seen that when $H$ and $(1+\omega_{eff})$ have the same sign, then $\dot{S_{h}}\geq 0$. Otherwise, $\dot{S_{h}}\leq 0$. Finally, in equation (\ref{sdott}), the sign of the total rate of change of the entropy, $\dot{S_{t}}$, is independent of the $\omega_{eff}$, while depending only on the sign of the hubble parameter. For an expanding/contracting universe, we have $\dot{S_{t}}\gtrless 0$.

In the following we will numerically investigate the validity of GSL on Chameleon model. In order to close the system of equations we make the following ansatz:
We take both potential and $f$ behave exponentially as $ V(\phi)=V_{0} e^{\delta_{1}\phi}$ and $ f(\phi)=f_{0} e^{\delta_{2}\phi}$ where $\delta_{1}, \delta_{2}, V_{0}$ and $f_{0}$ are arbitrary constants. There are no priori physical motivation for these choices, so it is only purely phenomenological which leads to the desired behavior of phantom crossing. The parameters $\delta_{1}$ and $\delta_{2}$ are dimensionless constants characterizing the slope of potential $V(\phi)$ and $f(\phi)$. Here, we assume that the universe is filled with cold dark matter, i.e. $\gamma=0$

The graphs in Fig.1 provide a dynamical universe with contraction for $t < 0$, bouncing at $t = 0$ and then expansion for $t > 0$. From equation (\ref{faydot}), one can see that for $\gamma=0$ the crossing occurs only for negative $f$. Also we see that at the bouncing point where the scale factor $a(t)$ is not zero we avoid singularity faced in the standard cosmology \cite{farajollahi}. In Fig.1a), we observe that the effective EoS parameter, $\omega_{eff}$ crosses the phantom  divide line twice at $t=0.75$ and $t=1.25$.\\

\begin{tabular*}{2.5 cm}{cc}
\includegraphics[scale=.27]{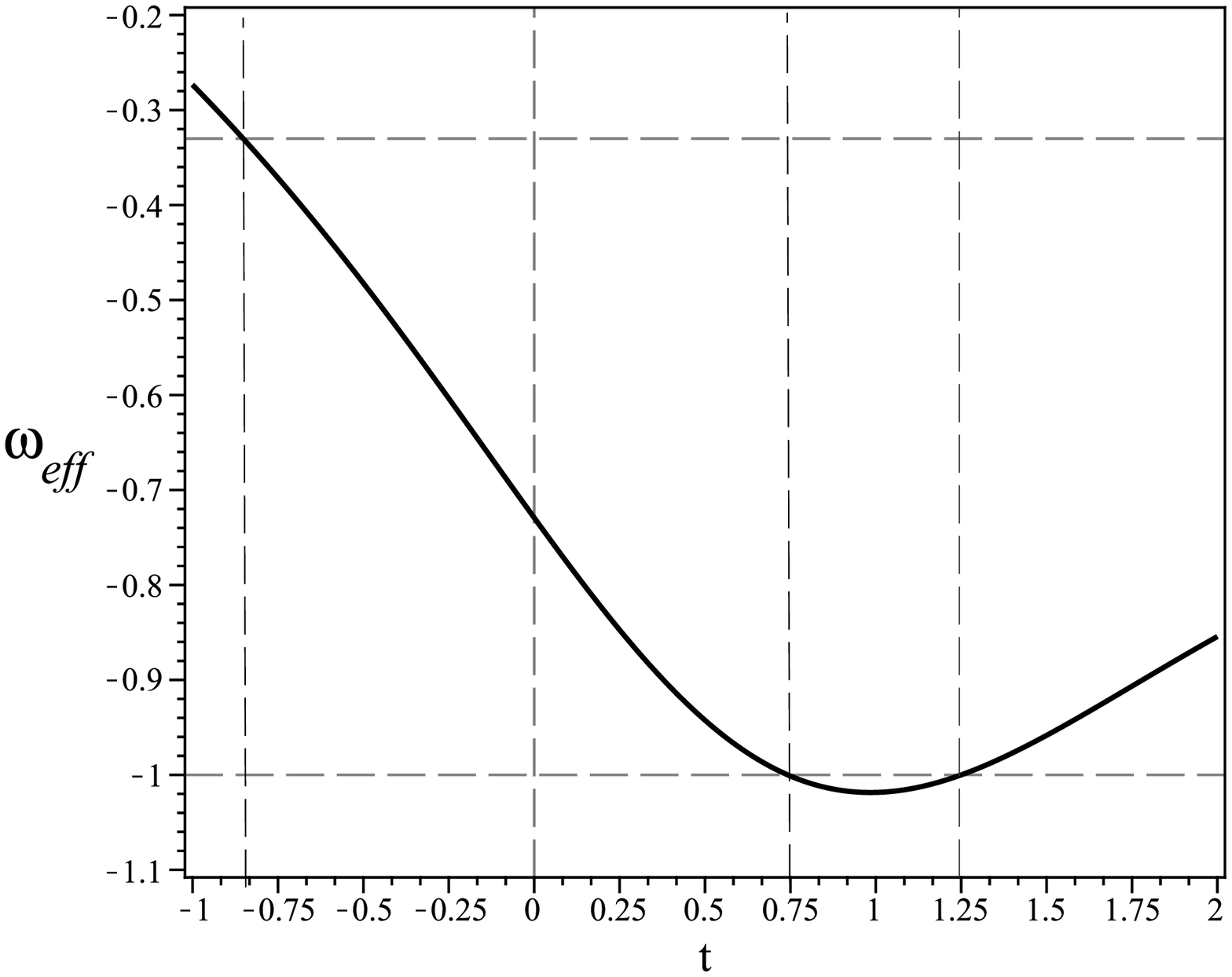}\hspace{0.5 cm}\includegraphics[scale=.27]{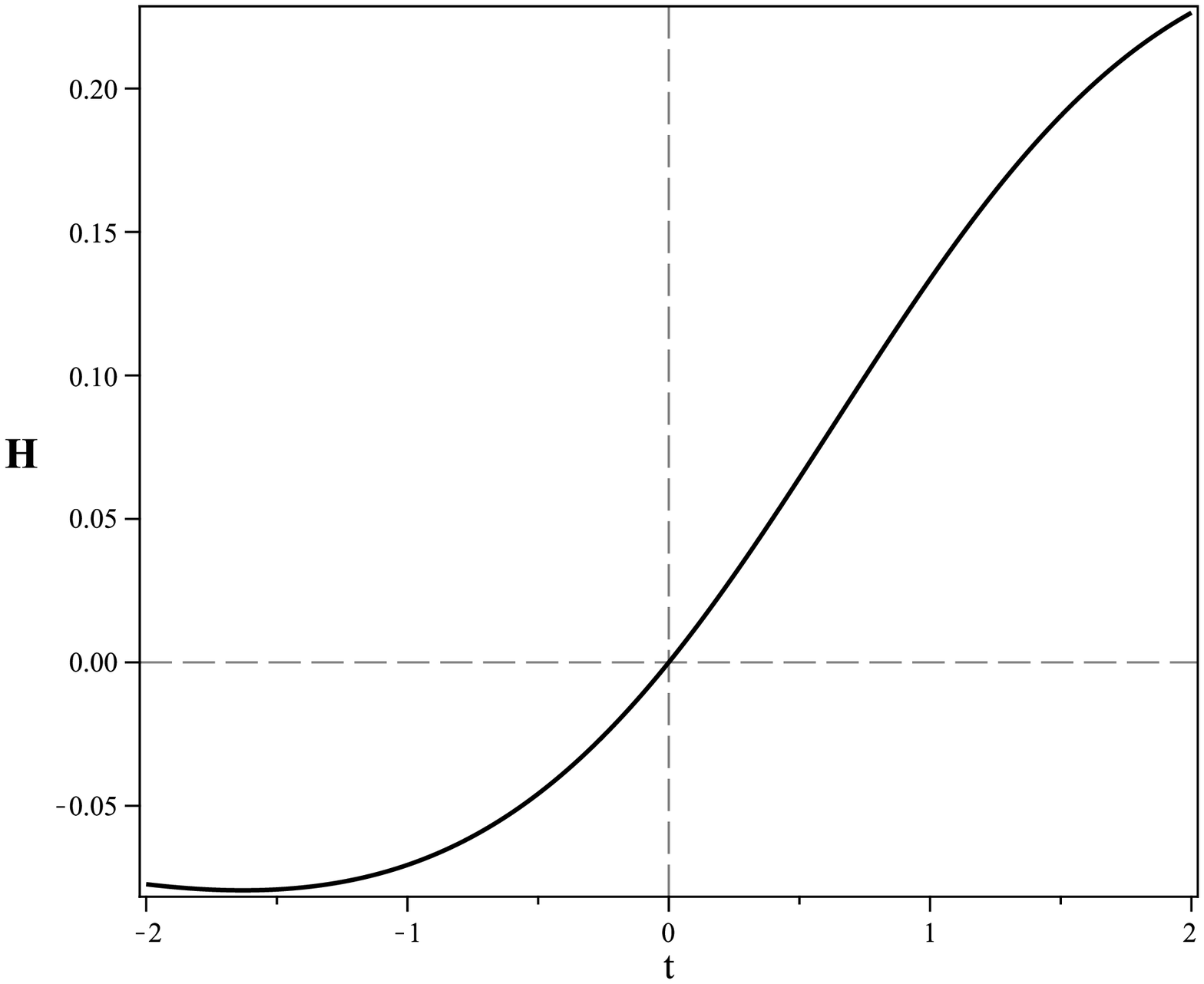}\hspace{0.5 cm}\includegraphics[scale=.27]{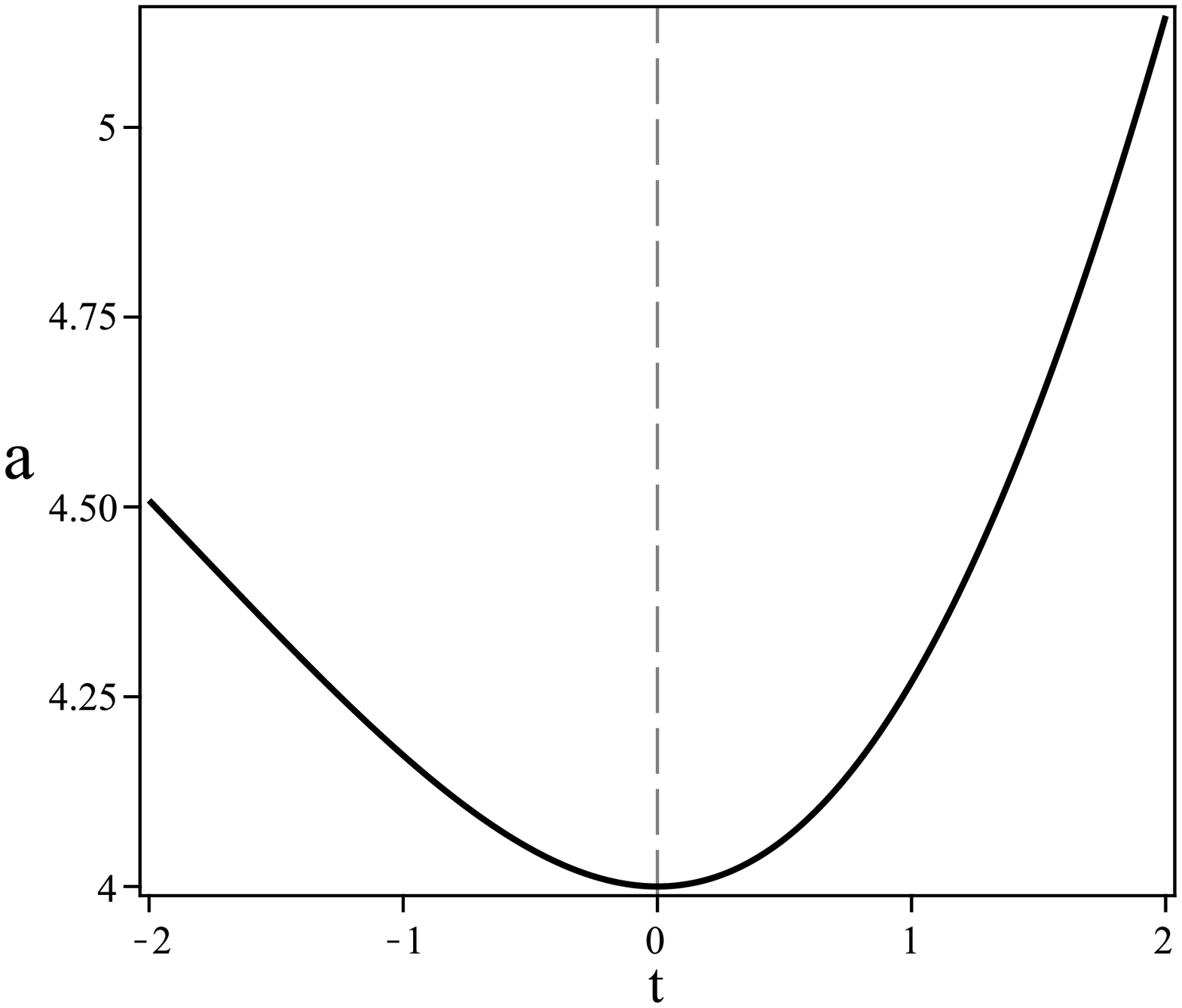}\\
\hspace{.5 cm}Fig. 1: The behavior of the effective EoS parameter, $\omega_{eff}$, hubble parameter, $H(t)$,\\ and the scale factor $a(t)$
with respect to the cosmological time $t$. The potential $V$ and $f$ are \\$V(\phi)=V_{0}\exp{(\delta_{1}\phi)}$ and $f(\phi)=f_{0}\exp{(\delta_{2}\phi)}$
where $f_{0}=-10$, $V_{0}=15$ and $\delta_{1}=-1$, $\delta_{2}=- 1$. \\ I.C. $\phi_{0}=1$, $\dot{\phi}(0)=-0.2$.\\
\end{tabular*}\\

In Fig.2) the rate of change of the internal entropy, horizon entropy and total entropy with respect to the effective EoS parameter in our model are shown. We also show the dynamics of the $H(t)$ for a comparison with the analytical findings. For example, for the internal entropy, from the graph, and for $H<0$, one observes that for $\omega_{eff}\geq-1/3$, then $\dot{S_{in}}\leq 0$; for  $-0.6\leq\omega_{eff}\leq-0.3$, then $\dot{S_{in}}\geq 0$; and for $-1\leq\omega_{eff}\leq-0.6$ when $H>0$ again $\dot{S_{in}}\leq 0$. Similar argument can be made for horizon and total entropies and as a result it shows that the numerical calculation is in complete agreement with the analytical findings.\\

\begin{tabular*}{2.5 cm}{cc}
\includegraphics[scale=.4]{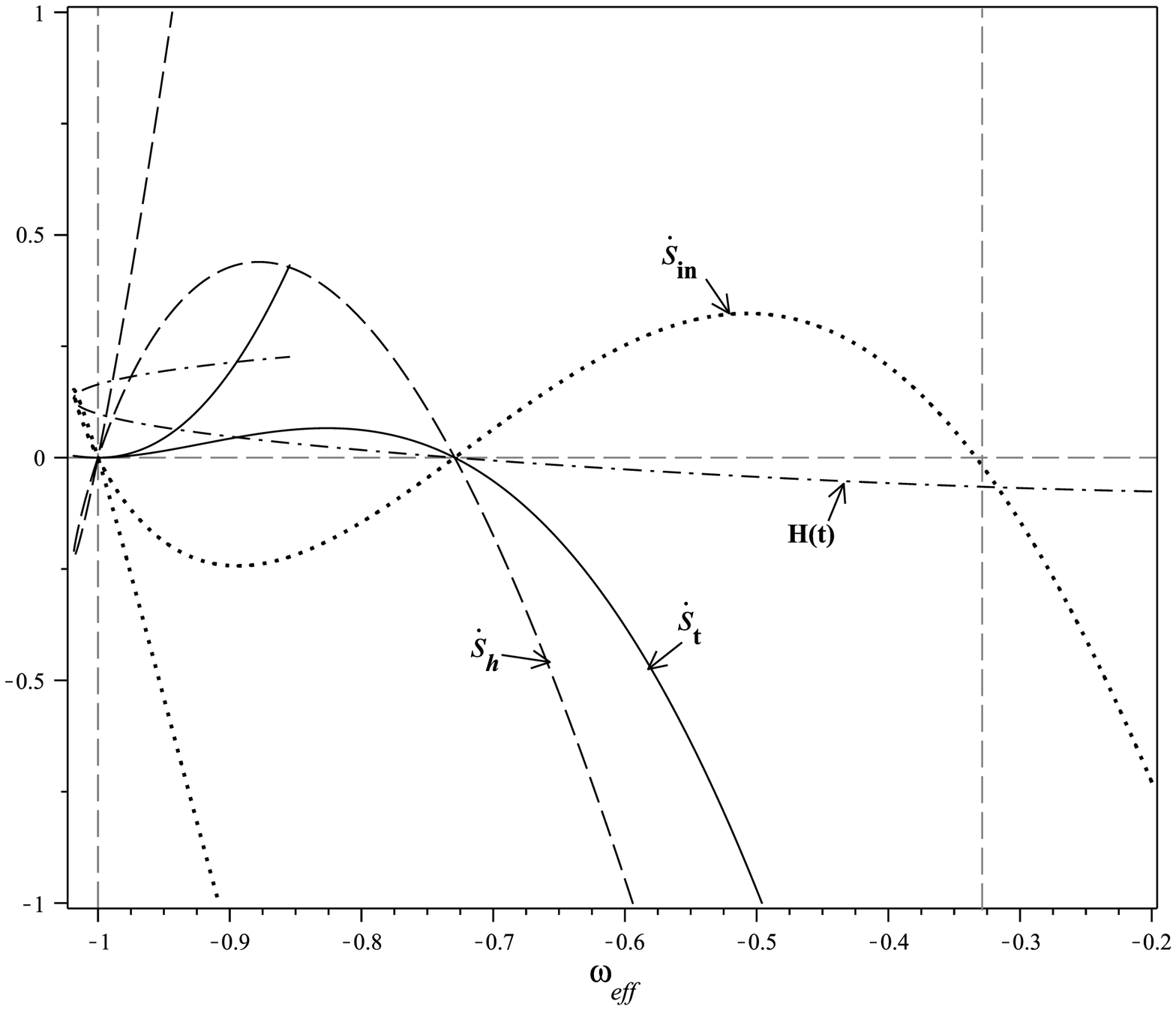}\\
\hspace{.5 cm}Fig. 2: \,The dynamics of the $\dot{S_{in}}$, $\dot{S_{h}}$, and $\dot{S_{t}}$ with respect to the effective EoS parameter, $\omega_{eff}$.\\
\end{tabular*}\\

In Fig.3, for $\omega_{eff}<-1$, we see that $\dot{S_{h}}<0$, whereas, both $\dot{S_{in}}$ and $\dot{S_{t}}$ are positive. The negative $\dot{S_{h}}$ is due to the phantom behavior of the dark energy. At $\omega_{eff}=-1$, as can be seen all three $\dot{S_{in}}$, $\dot{S_{h}}$, and $\dot{S_{t}}$ vanish, whereas, at $\omega_{eff}=-1/3$, only $\dot{S_{in}}$ vanishes, as expected.

\begin{tabular*}{2.5 cm}{cc}
\includegraphics[scale=.45]{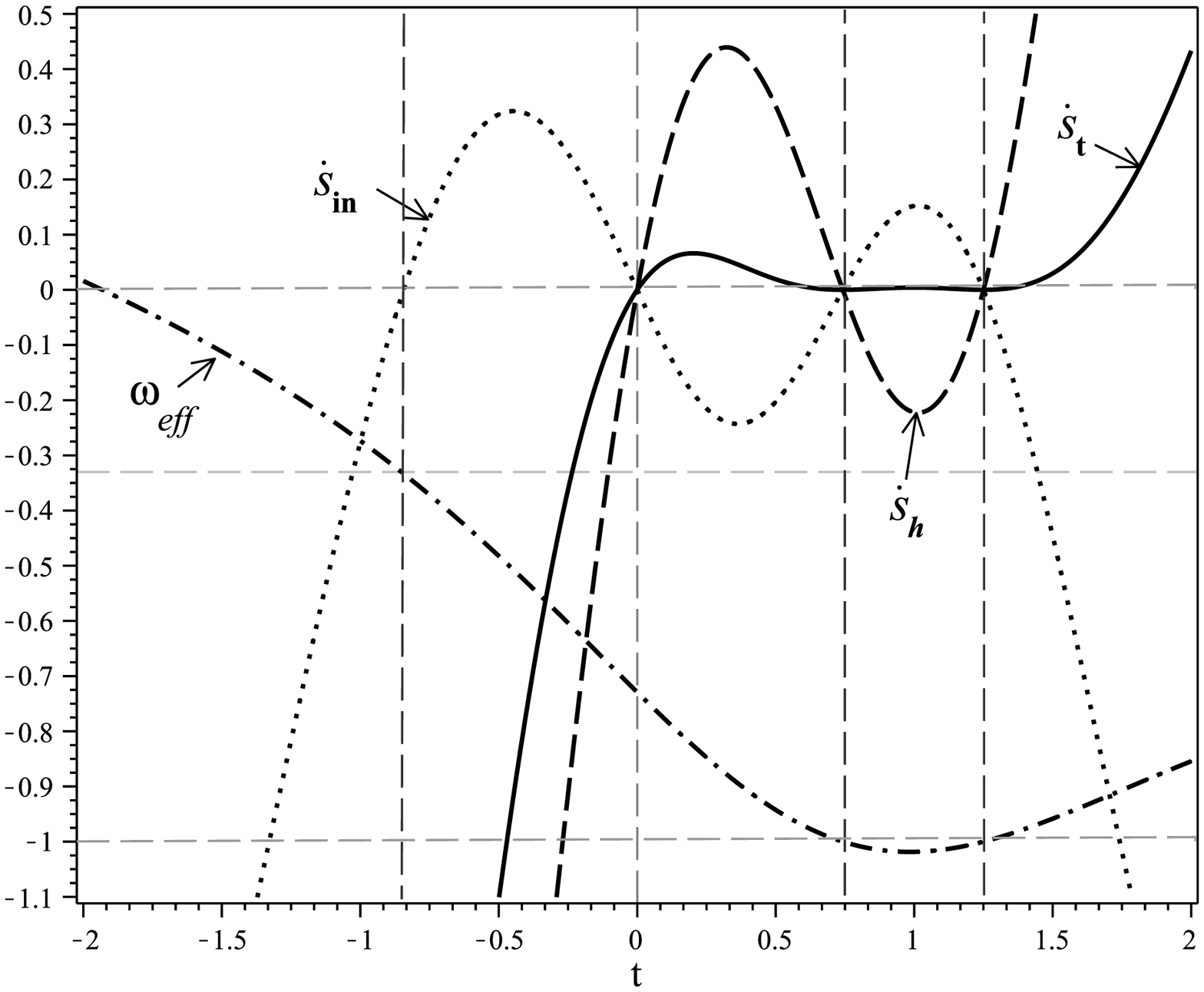}\\
\hspace{.5 cm}Fig. 3: The dynamics of $\dot{S_{in}}$, $\dot{S_{h}}$ and $\dot{S_{t}}$ with respect to the cosmological time \\
and in comparison to the dynamics of the effective EoS parameter, $\omega_{eff}$,\\
\end{tabular*}\\

Fig.4 shows that for $\gamma=0$ the crossing occurs only when $\dot{\phi^2}$ and $-\rho_{m}f$ are equal and at $t=t_{cross}=0.75, 1.25$,  $\dot{S_{t}}=0$. Also when $H=0$ at $t=0$, again $\dot{S_{t}}$ vanishes.\\

\begin{tabular*}{2.5 cm}{cc}
\includegraphics[scale=.5]{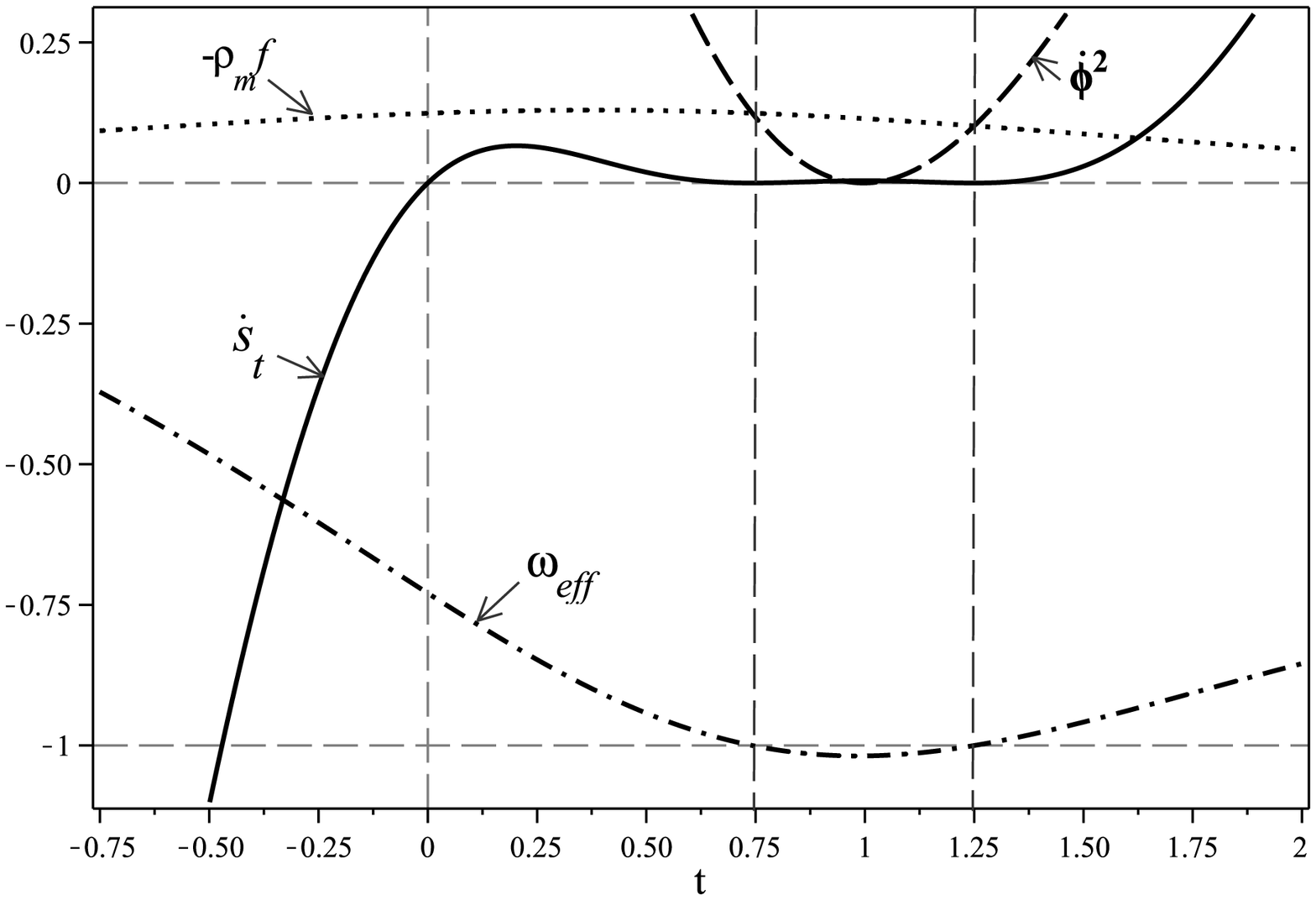}\\
\hspace{.5 cm}Fig. 4: The dynamics of $\dot{S_{t}}$ with respect to the cosmological time in comparison\\ with the $\dot{\phi^{2}}$, $-\rho_{m}f$ and $\omega_{eff}$.\\
\end{tabular*}\\

\section{Conclusion}

In this paper, we study the thermodynamics of the chameleon cosmological model in which a light scalar field (chameleon field) nonminimally coupled to the matter Lagrangian. We find that the evolution of the scale factor of the universe is non-singular in a bouncing cosmology, with an initial contracting phase which lasts until to a non-vanishing
minimal radius is reached and then smoothly transits into an expanding phase which provides a possible solution
to the singularity problem of standard Big Bang cosmology. We also show that the the total entropy of the universe in the model with bouncing behavior and dynamical effective EoS parameter increases/decreases with time in the expansion/contraction period. However,for the internal and horizon entropies, the sign of $\dot{S_{in}}$ and  $\dot{S_{h}}$ depend on the dynamics of both hubble and effective EoS parameters in the model. Further, even in an expanding universe, when $-1 <\omega_{eff}<-1/3$, the internal entropy decreases with time while the apparent horizon and total entropies do not. Another result we obtained  is that in an expanding universe in phantom era where $\omega_{eff}<-1$, the horizon entropy decreases with time which is due to the phantom behavior of the dark energy. Finally, we would like to mention that the thermodynamics of the bouncing universe with nonminimal coupling is an interesting topic. From a thermodynamical point of view, due to the modification of the theory of gravity in the early universe and its effects to the universe evolution, the subject deserves more attention in both theoretical and observational aspects.

\end{document}